# Konnektor: Connection Protocol for Ensuring Peer Uniqueness in Decentralized P2P Networks

By ONUR ÖZKAN
<contact@onurozkan.dev>

**ABSTRACT** Konnektor is a connection protocol designed to solve the challenge of managing unique peers within distributed peer-to-peer networks. By prioritizing network integrity and efficiency, Konnektor offers a comprehensive solution that safeguards against the spread of duplicate peers while optimizing resource utilization. This paper provides a detailed explanation of the protocol's key components, including peer addressing, connection initialization, detecting peer duplications and mitigation strategies against potential security threats.

## 1) INTRODUCTION

In decentralized networks, ensuring the uniqueness of each peer poses significant challenges. Let's consider an online multiplayer game. As the each peer corresponds to a user/player, maintaining the uniqueness of network peers becomes crucial for the game. Unlike centralized architectures where ensuring uniqueness is straightforward through querying centralized storage, decentralized architechtures lack such simplicity. This is because in decentralized networks each joining peer becomes part of the decision-making mechanism.

In order to keep a network peers distinct and unique, every new peer trying to join needs approval from others already in the network. However, if one peer joins and then refuses all other connection attempts, it can stop the network from growing. Furthermore, since ensuring uniqueness requires conducting network-wide checks when peers attempt to join, if a peer simultaneously makes multiple joining attempts, it can put pressure on the network's hardware resources and slow down overall operations performance.

Konnektor protocol is designed to solve these and numerous other problems. It requires peers to verify their identities with signed messages before joining the network. Connection establishment attempts are handled asynchronously and the connecting peer is tasked with performing CPU-intensive operations for a set duration, thus safeguarding the network against brute force attacks. Moreover, to prevent malicious peers from unconditionally rejecting all connection attempts, Konnektor requires peers claiming existing connections to the network to prove their authenticity through signed messages from the already connected peers.

# 2) PROTOCOL DESCRIPTION AND DETAILS

The design of Konnektor protocol includes digital signature, ConnectionBook, Entrypoint, and the following events:

**ConnectionInit**: Sent by peer (X) when it wants to connect another peer (Y) for the first time.

**AlreadyConnected**: Sent by peers claiming that they are already connected with a signature proving the claimed peer is indeed connected.

**NewPeer**: Peers receiving **ConnectionInit** event notify the network after certain validations by sending **NewPeer** event.

**ConnectionRequirement**: Peers receiving the **ConnectionInit** request increase the cost of this operation by using a proof-of-work algorithm, which can essentially be any type of proof-of-work function (such as the SHA256-based one used in Bitcoin) as desired by the implementer of this protocol. This ensures that peers attempting to spam or abuse **ConnectionInit** will likely inflict more harm upon themselves than upon the target peers.

**ConnectionRequirementResponse**: Peers receiving the **ConnectionRequirement** event respond with this event after performing the expected computing task.

**KeepAlive**: Peers send this event to each other at regular intervals to detect disconnections and timeouts. This can also be used as proof in the **AlreadyConnected** event.

Through the integration of these elements, Konnektor aims to guarantee the network's uniqueness and integrity.

## 2.1) Data Signing (Digital Signatures)

Konnektor protocol uses digital signatures in many stages to determine whether peers are impersonating other peers and whether the data they send is generated by themselves.

Any algorithm (preferably Ed25519 for speed and efficiency) that generates digital signatures and can be verified by public keys can be used with the Konnektor protocol. Konnektor uses public keys to define peer addresses, so when peers receive events they can verify the signature simply by using the sender address.

## 2.2) ConnectionBook

ConnectionBook is a thread-safe key-value structure for Konnektor protocol designed to securely manage information about connected peers. Using this peers keep track of events sent to each other to determine which peers they are connected to or in the process of connecting.

It utilizes timestamps for expiration to detect peer disconnections and timeouts. With each read/write operation, it checks the timestamp values of all registered peers and removes outdated entries before processing any operations. This ensures that all operations are performed on the most up-to-date data.

For each peer, ConnectionBook stores informations in the following format:

- **Key**: Address of the peer.
- **Value**:
    - **status**: Represents the current connection status, which can be one of the following:
        - *"Connected"*: Indicates that this peer is currently connected to us.
        - *"WantsToConnect"*: Used when peers initiated connection requests (with "ConnectionInit" event) to us, transitioning to "Connected" upon successful validation.
        - *"Connecting"*: Used when we are initiating connection request (with "ConnectionInit" event), transitioning to "Connected" upon receipt of a KeepAlive event (which is sent after successful validation from the target peer).
    - **last_event**: Contains the complete payload of the incoming request, including request signature and timestamp. Refer to chapter 2.3 for additional context.
    - **connected_at** (null if status isn't "Connected"): Timestamp indicating the peer's connected time.

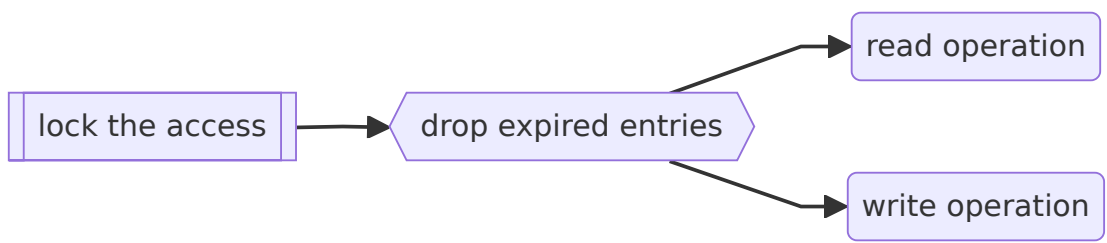

**Fig. 1:** Overview of the execution flow of each **ConnectionBook** operation.

For example, if we assume that peer X receives an event from peer Y, the operation on ConnectionBook for peer X would be as follows:

0. **Access Lock & Expired Entries**: Before proceeding with any operation, it's imporant to lock the access to the **ConnectionBook** to prevent data race conditions. This ensures that multiple threads or processes do not interfere with each other on read/write operations. After obtaining the lock, scan all entries and remove outdated ones by checking the timestamp field within the **last_event** field based on predefined/configured timeout limits for each status.
1. **Registration Check**:  Check if peer Y is registered in ConnectionBook.

2. **Upsert Entry**: If peer Y is registered, update last_event; if not, create a new entry.

## 2.3) The Entrypoint

This is the entry point of Konnektor protocol. Two things are checked in the entrypoint. Firstly, the integrity of the data with the sending peer is ensured, meaning it verifies that the data was indeed generated and sent by the sending peer. Secondly, it mitigates potential attacks (especially against resource attacks) using a rate limiter.

Each incoming event/request to the Entrypoint must stick to the following structure:

- **payload**: the actual payload of the request
  *"actual payload" means the data intended to be sent in the request is stored within this field. The root object contains only the timestamp, signature, and payload; it does not include any other data.*

- **timestamp**: UNIX timestamp in milliseconds
  *This value is used to determine that incoming requests are recently generated valid requests.*

- **signature**: the signed version of timestamp and payload (this value is used to determine that incoming requests are generated from the sender and not copied from other peers)
  *This value is used to determine that incoming requests are generated from the sender and not copied from other peers.*

Requests that do not comply with this format will be rejected from Entrypoint.

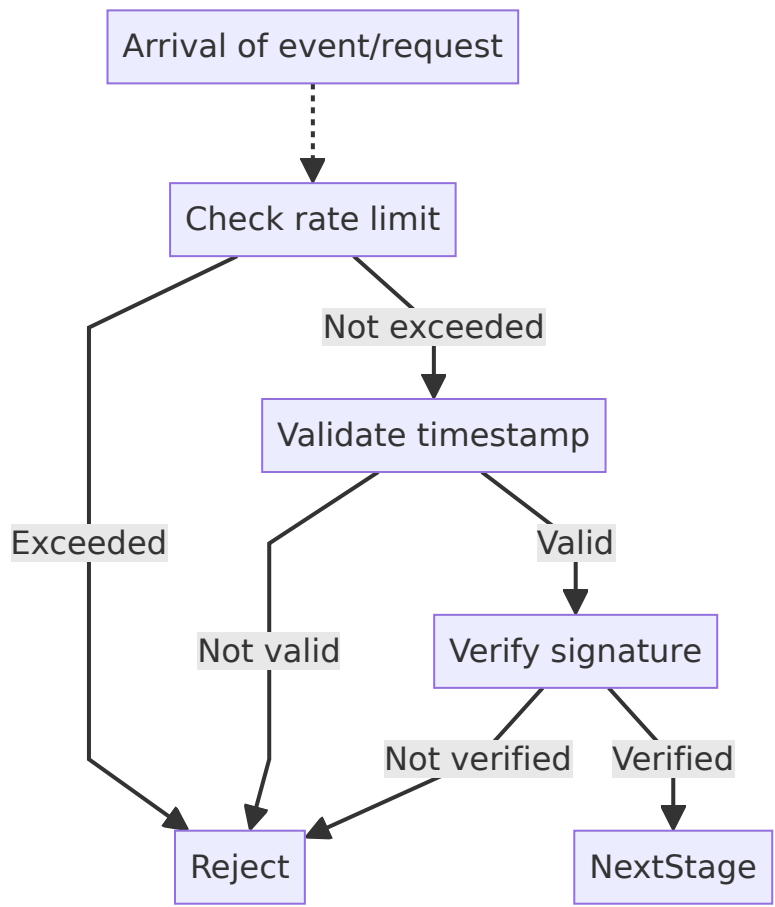

**Fig. 2:** Overview of how **Entrypoint** validates events.

1. **Arrival of Event/Request**: Incoming data/request is received by the **Entrypoint**.

2. **Rate-Limit**: Using a rate-limiting algorithm (various methods can be preferred here; for example, expirable hashmaps can be a suitable choice in terms of both effectiveness and lightweightness), it checks whether the sending peer has reached the rate limit, and if so, the request is rejected. The limit can be dynamically managed depending on the peer status (e.g., smaller limit can be used for peers that are not in the **ConnectionBook** with "Connected" status).

3. **Timestamp Validation**: Checks if the timestamp within the request is valid based on predefined/configured limitation. In other words, we verify that the timestamp is neither in the future nor in the past. If it's not valid, the request is rejected.

    *The reason* *for doing this is that we don't want peers to repeatedly use one signed payload indefinitely. This mechanism ensures signatures will expire at some point based on the predefined/configured limit on the receiver side.*

4. **Signature Verification**: The signature within the request is checked. If the signature hash doesn't match the address of the peer, timestamp and payload of the request that means the peer is trying to impersonate another peer. If that's the case, request is rejected.

If the request successfully passes all the checks outlined above, it proceeds to the next stage.

## 2.4) Sending ConnectionInit: Initiating Connections with Other Peers

This is the initial stage of the Konnektor protocol for establishing connections between peers.

Each **ConnectionInit** request contains the following fields:

- **payload**:
    - **target_peers**: Addresses of the peers to connect to.
        - **timestamp**: Current UNIX timestamp in milliseconds.
        - **signature**: The signed version of the **timestamp** and **payload**.

    *Please note that we only require the **target_peers** field for the **ConnectionInit** event. However, **timestamp**, **signature**, and **payload** fields are necessary for each request for **Entrypoint** validations.*

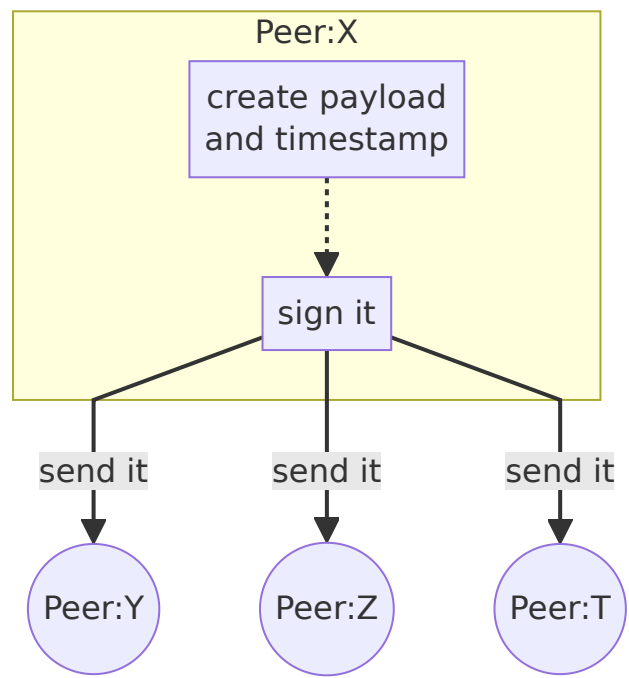

**Fig. 3:** Overview of how a peer initiates connection establishment with other peers.

When a peer wishes to connect to a network, they must follow these steps:

1. **Preparing ConnectionInit Payload**: Initially, a payload for the ConnectionInit request is generated
2. **Sending ConnectionInit**: Once the payload for ConnectionInit is generated according to the above format, it is then sent to all peers within the **target_peers** and each of these addresses will be registered into the **ConnectionBook** with the "Connecting" status.

With the completion of the above steps, the initial connection phase for peers listed in the **target_peers** is finalized. These peers are expected to verify the received **ConnectionInit** event and respond by sending back **ConnectionRequirement** event (refer to chapter 2.5 to see how it's done).

## 2.5) Receiving ConnectionInit Events

Peers understand that there is a peer attempting to connect to them upon receiving the ConnectionInit event. In this case, these peers perform certain checks to verify the validity of the sender peer and payload in the incoming request.

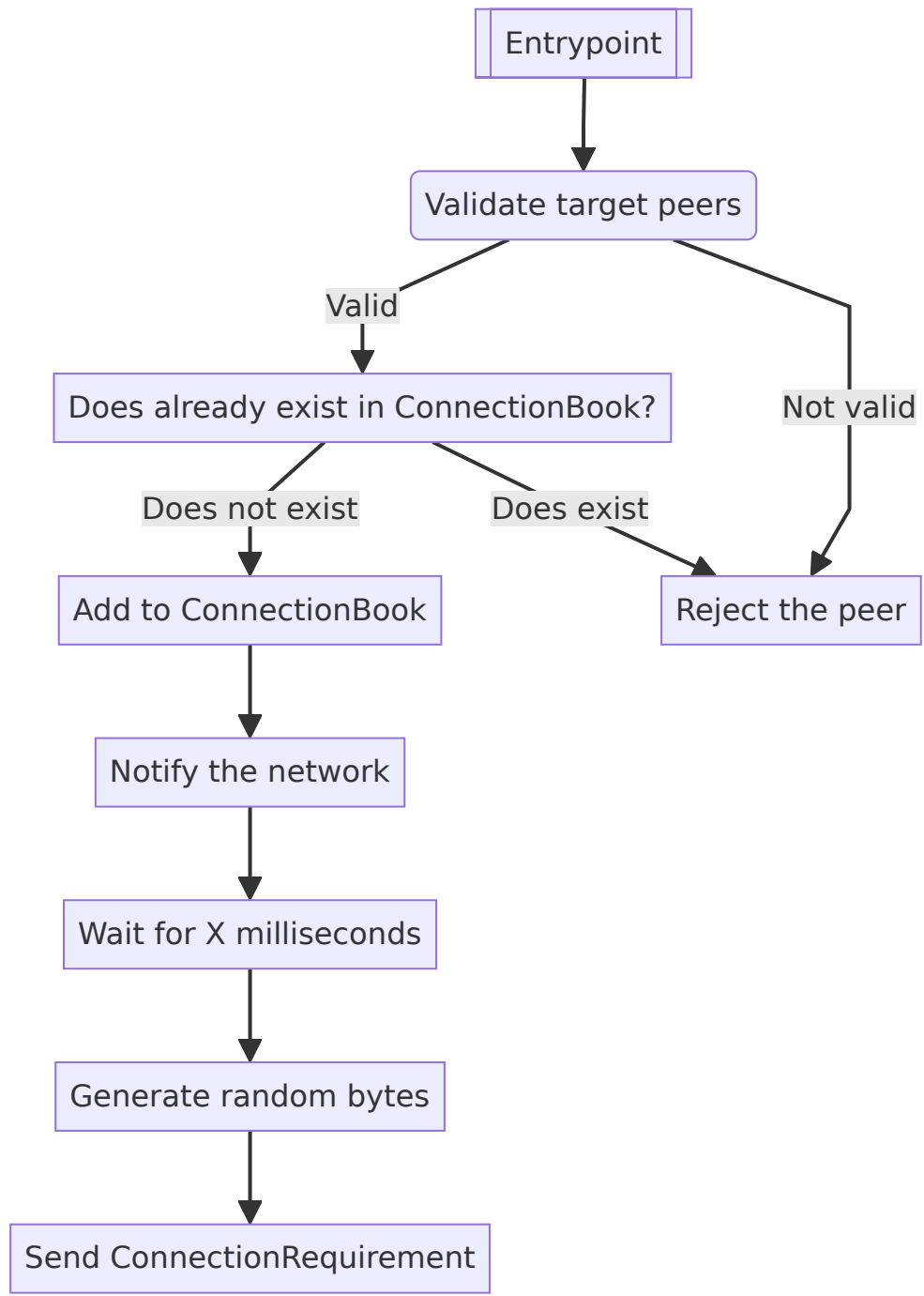

**Fig. 4:** Overview of how a peer handles a **ConnectionInit** event upon receiving it.

0. **Entrypoint**
1. **Validating Target Peers**: If our address is not included in the **target_peers** of the **ConnectionInit** payload, it indicates an invalid request and the process ends here (simply by ignoring it).

2. **Check for Existing Peer in ConnectionBook**: If the requesting peer already exists in the **ConnectionBook**, the connection request is rejected, and an **AlreadyConnected** (chapter 2.9) event is propagated to the entire network.

    *This prevents duplicate connection attempts from the same peer and maintains uniqueness within the network.*

3. **Add Requesting Peer to ConnectionBook**: If the requesting peer does not exist in the ConnectionBook, it is added into the **ConnectionBook** along with the **ConnectionInit** payload with the "WantsToConnect" status.

4. **NewPeer Event: Notify the Network**: Propagate **NewPeer** event to the network with the following payload:

- **payload**: entire/raw payload of the received *ConnectionInit* event
- **timestamp**: Current UNIX timestamp in milliseconds.
- **signature**: The signed version of the **timestamp** and **payload**.

*Please note that we only care about the raw payload of **ConnectionInit** event. However, **timestamp**, **signature**, and **payload** fields are necessary for each request for **Entrypoint** validations.*

5. **Handling Peers Asynchronously**: Select a random value from a predefined/configured range (e.g., 0-5000ms) and wait for that many milliseconds.

    *As explained in the step 5, peers receiving **ConnectionInit** event requests generates random bytes to requesting peers for signing. Receiving too many **ConnectionInit** requests from different peers would lead to constant generation of random bytes causing resource exhaustion. We avoid this problem by generating random bytes only once at unpredicted time and send it to the all peers registered in the ConnectionBook. Instead of generating random bytes for each **ConnectionInit** sender peer, we generate the random bytes once every X milliseconds (where X is randomly chosen from the predefined/configured range) for all the peers with "WantsToConnect" waiting in the **ConnectionBook**.*

6. **Random Byte Generation**: Generate random bytes of the specified size as predefined/configured and UNIX timestamp (for timeout).

    *The main reason for generating and signing a random byte array and including a UNIX timestamp is to increase the space and time complexity for peers attempting brute-force attacks using the ConnectionInit events. Increasing the brute-force cost for the attacker, coupled with rate limiting on victim peers significantly minimizes the impact of brute-force/resource attacks.*

7. **Sending ConnectionRequirement**: Send a **ConnectionRequirement** event request with the generated byte array, difficulty and signature hash to all the peers with the "WantsToConnect" status in the **ConnectionBook**.

## 2.6) Receiving NewPeer Events

When a peer receives a connection request, it tells the entire network about the new peer by sending out a **NewPeer** event. This helps to identify and reject any previously connected peers.

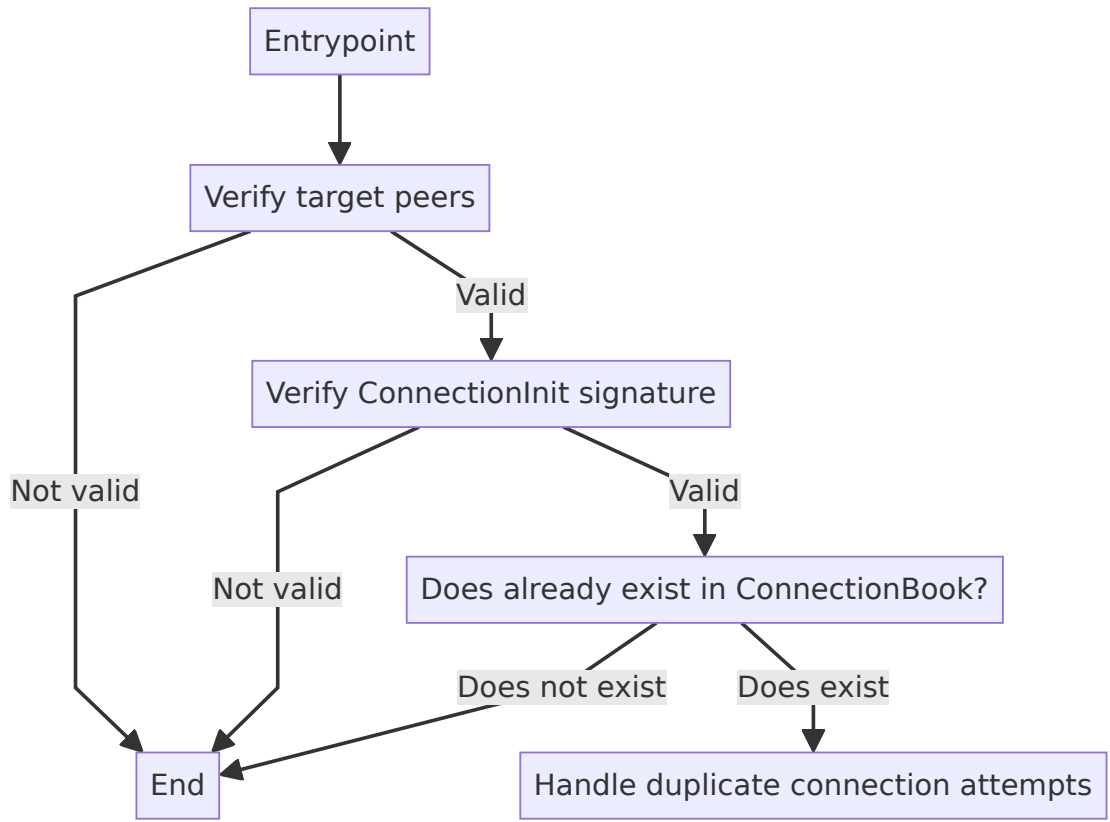

**Fig. 5:** Overview of how a peer handles a **NewPeer** event upon receiving it.

Each peer that receive the "NewPeer" event in the network responsible for performing the following checks:

0. **Entrypoint**

1. **Verification of target_peers**: First, they check if the peer that sent the event is listed in the **target_peers** field within the **ConnectionInit** event, which is part of the payload of the **NewPeer** event. If it's not listed there, consider the event invalid and simply ignore it.

2. **Signature Verification of ConnectionInit**: If the sending peer is indeed listed in the **target_peers**, the receiving peer then verifies the signature (not the signature of **NewPeer** event, which is already verified from **Entrypoint**) of the **ConnectionInit** event. If the signature doesn't match or is invalid, consider the event invalid and ignore it.

3. **ConnectionBook Check**: Assuming the event passes all the checks above, the receiving peer then looks to see if the peer trying to establish a connection already exists in the **ConnectionBook**. If the peer isn't found there, it indicates that there's no duplicate with us and the **NewPeer** validation ends here.

4. **Handling Duplicate Connection Attempts**: However, if the peer does exist in the **ConnectionBook**, if **target_peers** of the payload and **target_peers** of the **ConnectionBook** don't match (which means this peer trying to connect again which already have connected to this network), we send an **AlreadyConnected** event to the network. This event includes the last event sent by this peer stored in the **last_event** field in the **ConnectionBook**. **last_event** will contain a recently signed message from this peer, proving that this peer is indeed already connected to this network, as we have the signature and message from it.

## 2.7) Receiving ConnectionRequirement Events

When peer sends a **ConnectionInit** to the peers it wants to connect with, it promptly receives a **ConnectionRequirement** event. This event triggers a hashing process on the requesting peers. This hashing process raises the difficulty and cost of establishing a connection to the network, making it more challenging for attackers to carry out brute-force attacks effectively. This effect is further enhanced when combined with rate-limiting mechanisms on the receiver side.

Peers receiving the ConnectionRequirement event must complete the following steps:

0. **Entrypoint**
1. **Prove Work**: Compute the hash of given payload (the byte array generated from other peer) which aligns with the expected difficulty.
2. **Sending ConnectionRequirementResponse**: Send the generated data as a **ConnectionRequirementResponse** event to the **ConnectionRequirement** sender peer.

The payload of the **ConnectionRequirementResponse** event includes:

- **payload**:
    - **requirement_raw_payload**: The raw payload of the received ConnectionRequirement event. This is used for cross-validation purposes from the receiver side.
    - **proof**: A hash, computed proof-of-work result.
- **timestamp**: Current UNIX timestamp in milliseconds.
- **signature**: The signed version of the **timestamp** and **payload**.

*Please note that we only care about the **requirement_raw_payload** and **proof** fields. However, **timestamp**, **signature**, and **payload** fields are necessary for each request for **Entrypoint** validations.*

## 2.8) Receiving ConnectionRequirementResponse Events

Peers receiving **ConnectionRequirement** event send a **ConnectionRequirementResponse** event after completing the expected computing process. At this stage, the computed hash/proof of the byte array and difficulty (from the **ConnectionRequirement** event) is checked for validity.

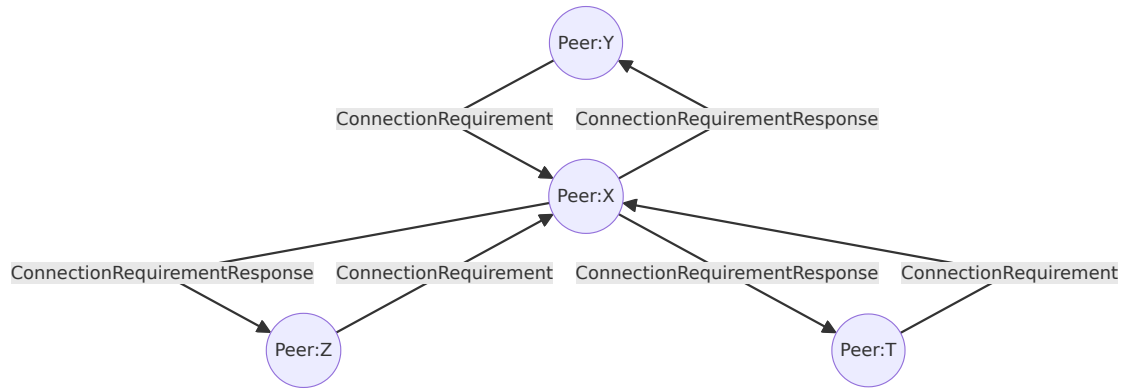

**Fig. 6:** A peer (X) who wants to connect to other peers (Y, Z, and T) receives a **ConnectionRequirement** event from each of them. It is the responsibility of peer X to send a **ConnectionRequirementResponse** back to each peer.

The handling process is as follows:

0. **Entrypoint**
1. **Timestamp Validation**: **timestamp** in the **requirement_raw_payload** field is checked and it is determined whether the proof-of-work has been performed on a new **ConnectionRequirement** within the configured limit (e.g., 30 seconds). If the **timestamp** is older than the limit, the connection request with the peer is rejected and it is removed from the **ConnectionBook**.
2. **Payload Verification**: The **signature** within the **requirement_raw_payload** is checked. This ensures that the byte array, difficulty and timestamp indeed belongs to us.
3. **Proof Validation**: Validate the proof by checking the hash value from **proof** field against the byte array and difficulty from the **requirement_raw_payload** field.
4. **Update of Peer Status**: If everything is valid, update peer's status in the **ConnectionBook** from "WantsToConnect" into "Connected", fill **connected_at** field and start sending **KeepAlive** events (for every x seconds which can be configurable) to this peer.

## 2.9) Receiving AlreadyConnected Events

When a peer attempts to reconnect to an already connected network, an "AlreadyConnected" event is broadcasted to the entire network by the peers already connected to it.

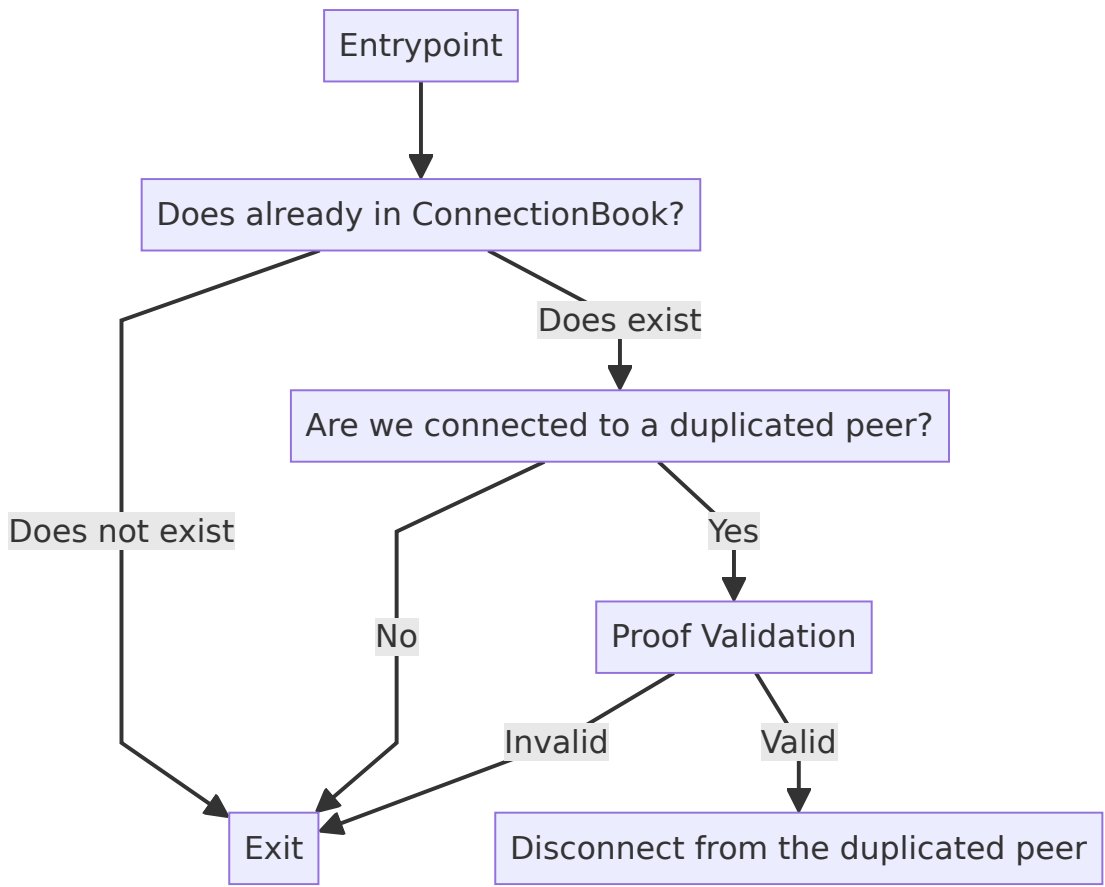

**Fig. 7:** Overview of how a peer handles a **AlreadyConnected** event upon receiving it.

Upon receiving this event, each peer applies the following steps:

0. **Entrypoint**

1. **ConnectionBook Check**: Check if the mentioned already connected peer is registered in the ConnectionBook. If not, no further action is required and the check is complete.

2. **Duplication Check**: The payload of the **AlreadyConnected** event contains a recently signed event from the peer who wishes to join this network. This event can be either a **ConnectionInit** or a **KeepAlive**. Both of these events contain a field called **target_peers**, indicating the list of peers this peer is attempting to connect to (if the event is a **ConnectionInit**) or already connected to (if the event is a **KeepAlive**). We check if the **target_peers** in the **ConnectionBook** is equal to the **target_peers** data within the **AlreadyConnected** payload.

    The duplication check should be handled as demonstrated below:

    ```
    # For KeepAlive events if target_peers don't match, it
    means we are connected/connecting to the duplicated peer.
     keep_alive.target_peers !=
    connection_book_last_event.target_peers

     # For ConnectionInit events, if target_peers match, it
    means wer are connected/connecting to the duplicated peer.
     connection_init.target_peers ==
    connection_book_last_event.target_peers
    ```

3. **Data Validation**: If duplication check returns true(meaning we are connected or connecting to the duplicated peer), verify the **signature** within the **AlreadyConnected** payload with the **target_peers**. If the signature is valid, remove the peer from the **ConnectionBook** and disconnect from it.

Through this flow duplicated peers are identified and their connections are terminated before the connection is fully established (at worst, very shortly after it begins), ensuring network uniquness and mitigating potential issues associated with duplicate connections.

## 2.10) Handling KeepAlive Events

**KeepAlive** events are used in timeout/disconnection mechanisms. Peers use these events to keep track of connected peers and maintain the **ConnectionBook**.

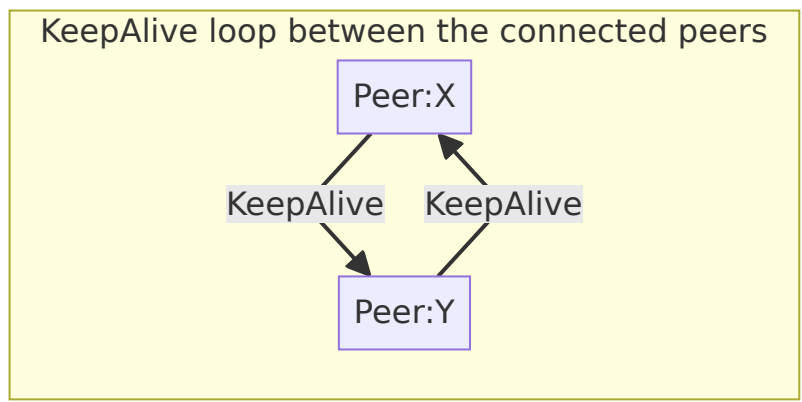

**Fig. 8:** Overview of the utilization of **KeepAlive** events between the connected peers.

Each **KeepAlive** event includes the following payload:

- **payload**:
    - **target_peers**: Target addresses of peers which this event was generated and going to be sent for.
- **timestamp**: Current UNIX timestamp in milliseconds.
- **signature**: The signed version of the **timestamp** and **payload**.

*Please note that we only require the **target_peers** field for the **ConnectionInit** event. However, **timestamp**, **signature**, and **payload** fields are necessary for each request for **Entrypoint** validations.*

Using the **timestamp** and the **signature** (that is also validated from the **Entrypoint**) alongside the **target_peers** values, **KeepAlive** events also serve as proof in **AlreadyConnected** events.

Peers receiving KeepAlive events execute the following steps:

0. **Entrypoint**

1. **ConnectionBook Check**: Check if the sender peer exists in the **ConnectionBook**. If it doesn't, it means this is an invalid event. No further action is required and the process ends here.

2. **Status Check**: If the sender peer exists in the **ConnectionBook**, check its status. If it's not "Connected" or "Connecting", it means this is an invalid event. No further action is required and the process ends here..

3. **Handling for 'Connecting' Peers**: If status of the peer in **ConnectionBook** was "Connecting", that means we were trying to connect to this peer and it has been accepted the connection (see chapter 2.8 Receiving ConnectionRequirementResponse Events). Switch the status from 'Connecing' to 'Connected', fill "connected_at" field and start sending **KeepAlive** events (for every x seconds which can be configurable) to this peer. Also, update the **last_event** value in the **ConnectionBook** with the raw payload of the received **KeepAlive** event.

4. **Handling for 'Connected' Peers** If status was "Connected" already, then just update the **last_event** value in the **ConnectionBook** with the raw payload of the received **KeepAlive** event.

## 3) Implementation Notes

Konnektor implementation should provide a configuration interface to users/developers to adjust various settings like rate limiting, connection timeouts, and payload size and hash difficulty for **ConnectionRequirements** for their peers. It is necessary to allow for such customization so that individuals can increase or decrease the various thresholds as they see fit.

Additionally, rate limiter in **Entrypoint** should be optional; allowing users to disable it if not required. This is especially important in setups where the rate limiter is managed separately (such as by a load balancer). Therefore, providing the option to turn it off when unnecessary can be useful for certain people.

## 4) Summary

In this paper we introduced Konnektor, the first connection protocol designed to ensure peer uniqueness in decentralized peer-to-peer networks. Through its components such as digital signatures, **ConnectionBook** and event-based interactions like **ConnectionInit** and **NewPeer**, Konnektor provides a practical solution for managing peer identities and connections. By utilizing specialized techniques like requiring connecting peers to perform computing tasks to increase joining costs, Konnektor effectively raises the barrier against malicious actors and resource attacks ensuring the integrity and stability of decentralized networks. Additionally, Konnektor's reliance on digital signatures for message authentication guarantees the authenticity of peer identities, enhancing trust and security within the network. Overall, Konnektor provides a complete solution for maintaining network uniqueness and security setting the stage for robust and resilient decentralized systems.